\documentclass{css2022}

\usepackage{epsfig}

\newcommand{\PACS}{\MSC}

\title{The Weak Field Approximation of General Relativity, Retardation, and the Problem of Precession of the Perihelion for Mercury}

\author{Asher Yahalom\\
\vskip 2mm {\small
Ariel University\\
Ariel 40700, Israel\\
asya@ariel.ac.il}}

\abstract{
In this paper we represent a different approach to the calculation of the perihelion shift than the one presented in common text books. We do not rely on the Schwarzschild metric and the Hamilton Jacobi technique to obtain our results. Instead we use a weak field approximation, with the advantage that we are not obliged to work with a definite static metric and can consider time dependent effects. Our results support the conclusion of K\v{r}\'{ı}\v{z}ek
\cite{Krizek} regarding the  significant influence of celestial parameters
on the indeterminacy of the perihelion shift of Mercury’s orbit. This shift is thought
to be one of the fundamental tests of the validity of the general theory of relativity. In the
current astrophysical community, it is generally accepted that the additional relativistic
perihelion shift of Mercury is the difference between its observed perihelion shift and the
one predicted by Newtonian mechanics, and that this difference equals $43"$ per century.
However, as it results from the subtraction of two inexact numbers of almost equal
magnitude, it is subject to cancellation errors. As such, the above accepted value is highly
uncertain and may not correspond to reality.}



\begin{document}

\maketitle

\newcommand{\beq} {\begin{equation}}
\newcommand{\enq} {\end{equation}}
\newcommand{\ber} {\begin {eqnarray}}
\newcommand{\enr} {\end {eqnarray}}
\newcommand{\eq} {equation}
\newcommand{\eqs} {equations }
\newcommand{\mn}  {{\mu \nu}}
\newcommand{\abp}  {{\alpha \beta}}
\newcommand{\ab}  {{\alpha \beta}}
\newcommand{\sn}  {{\sigma \nu}}
\newcommand{\rhm}  {{\rho \mu}}
\newcommand{\sr}  {{\sigma \rho}}
\newcommand{\bh}  {{\bar h}}
\newcommand{\br}  {{\bar r}}
\newcommand {\er}[1] {equation (\ref{#1}) }
\newcommand {\ern}[1] {equation (\ref{#1})}
\newcommand {\Ern}[1] {Equation (\ref{#1})}
\newcommand{\hdz}  {\frac{1}{2} \Delta z}

\section {Introduction}

Under Newtonian physics, an object in an (isolated) two-body system, consisting of the object orbiting a spherical mass, would trace out an ellipse with the massive object at a focus of the ellipse. The point of closest approach, called the periapsis (or, because the central massive body in the Solar System is the Sun, perihelion), is fixed. Hence the major axis of the ellipse remains fixed in space. Both objects orbit around the center of mass of this system, so they each have their own ellipse, but the heavier body trajectory is smaller. In fact it can
be much smaller if the ratio between the masses is considerable. However, a number of effects in the Solar System cause the perihelia of planets to precess (rotate) around the Sun, or equivalently, cause the major axis to rotate, hence changing its orientation in space. The principal cause is the presence of other planets which perturb one another's orbit. Another (much less significant) effect is solar oblateness.

Mercury deviates from the precession predicted from these Newtonian effects. This anomalous rate of precession of the perihelion of Mercury's orbit was first recognized in 1859 as a problem in celestial mechanics, by Urbain Le Verrier. His re-analysis of available timed observations of transits of Mercury over the Sun's disk from 1697 to 1848 demonstrated that the actual rate of the precession disagreed from that predicted from Newton's theory by $38"$ (arcseconds) per tropical century \cite{Verrier} (later  it was estimated to be $43"$ by Simon Newcomb in 1882 \cite{Newcomb}). A number of ad hoc and ultimately unsuccessful solutions were proposed, but they seemed to cause more problems. Le Verrier speculated that another hypothetical planet might exist to account for Mercury's behavior \cite{Levenson}. The previously successful search for Neptune based on its perturbations of the orbit of Uranus led astronomers to place some faith in this possible explanation, and the hypothetical planet was even named Vulcan. Finally, in 1908, W. W. Campbell, Director of the Lick Observatory, after meticulous photographic observations by Lick astronomer, Charles D. Perrine, at three different solar eclipse expeditions, stated, "In my opinion, Dr. Perrine’s work at the three eclipses of 1901, 1905, and 1908 brings the observational side of the famous intramercurial-planet problem definitely to a close." \cite{Baum}. Since no evidence of Vulcan was found and Einstein's 1915 general theory accounted for Mercury's anomalous precession. Einstein could write to his friend Michael Besso, "Perihelion motions explained quantitatively...you will be astonished" \cite{Baum}.

In general relativity, this remaining precession, or change of orientation of the orbital ellipse within its orbital plane, is explained by gravitation being mediated by the curvature of spacetime,
and by the fact that the trajectory must be a geodesic in the curved space-time. Einstein showed that general relativity \cite{Ein,Einstein2} agrees closely with the observed amount of perihelion shift. This was a powerful factor motivating the adoption of general relativity.

Although earlier measurements of planetary orbits were made using conventional telescopes, more precise measurements are now made using a radar. The total observed precession of Mercury is $574.10" \pm 0.65$ per century \cite{Clemence} relative to the inertial International Celestial Reference System (ICRS) (the current standard celestial reference system adopted by the International Astronomical Union (IAU). Its origin is at the barycenter of the Solar System, with axes that are intended to be oriented with respect to the stars.)

This precession can be attributed to the causes \cite{Clemence,Park} described in table \ref{prehelexth}.
\begin{table}
  \centering
\begin{tabular}{|c|c|}
  \hline \hline
  \parbox[t]{5.2cm}{\bf Causes of the precession of perihelion for Mercury
 (arcsec/Julian century)} & {\bf Cause} \\ \hline \hline
  $532.3035$ & Gravitational tugs of other solar bodies \\ \hline
  $0.0286$  & Oblateness of the Sun (quadruple moment) \\ \hline
  $42.9799$ &  General Relativity effect (Schwarzschild - like) \\ \hline
  $-0.0020$ & Lense-Thirring precession \\ \hline \hline
  ${\bf 575.31}$ &  {\bf Total predicted} \\ \hline \hline
  ${\bf 574.10 \pm 0.65}$ & {\bf Observed} \\
  \hline \hline
\end{tabular}
  \caption{The different contributions to the precession of perihelion for Mercury
amount (arcsec/Julian century), theory vs. prediction.}
  \label{prehelexth}
\end{table}
Thus despite the efforts, theoretical predictions of the precession of perihelion for Mercury do not full within observational results error bars and the discrepancy is at best $0.56"$ per century or about $0.1 \%$. This is not much but still requires explanation.

In general relativity the perihelion shift $\delta \theta$, expressed in radians per revolution, is approximately given by \cite{padma}:
\beq
\delta \theta = \frac {6 \pi G M}{a c^{2}\left(1-e^{2}\right)}
\label{prehprec}
\enq
$G \simeq 6.67 \ 10^{-11} \ {\rm m}^3 {\rm kg}^{-1} {\rm s}^{-2}$ is the universal gravitational constant and $c \simeq 3 \ 10^8  \ {\rm m} {\rm s}^{-1}$ indicates the velocity of light in the absence of matter, $M$ is the mass of the sun, $a$ is the semi-major axis of the ellipsoidal trajectory and $e$ is its orbital eccentricity. The above is based on calculating a geodesic in a Schwarzschild geometry, that is in a geometry created by a static point mass. The framework of Schwarzschild geometry does not allow us to take into account effects such as the motion of the sun with respect to the frame.

The other planets experience perihelion shifts as well, but, since they are farther from the Sun and have longer periods, their shifts are lower, and could not be observed accurately until long after Mercury's. For example, the perihelion shift of Earth's orbit due to general relativity is theoretically $3.83868"$ per century and experimentally $3.8387\pm0.0004"$/cy, Venus's is $8.62473"$/cy and $8.6247 \pm 0.0005"$/cy. Both values have now been measured, with results in good agreement with theory \cite{biswa}.

Einstein's general relativity (GR) is known to be invariant under general coordinate modifications.
This group of general transformations has a Lorentz subgroup, which is valid even in the weak field approximation. This is seen through the field equations containing the d'Alembert (wave) operator, which can be solved using a retarded potential~solution.

\begin{figure}
\centering
\includegraphics[width= 11 cm]{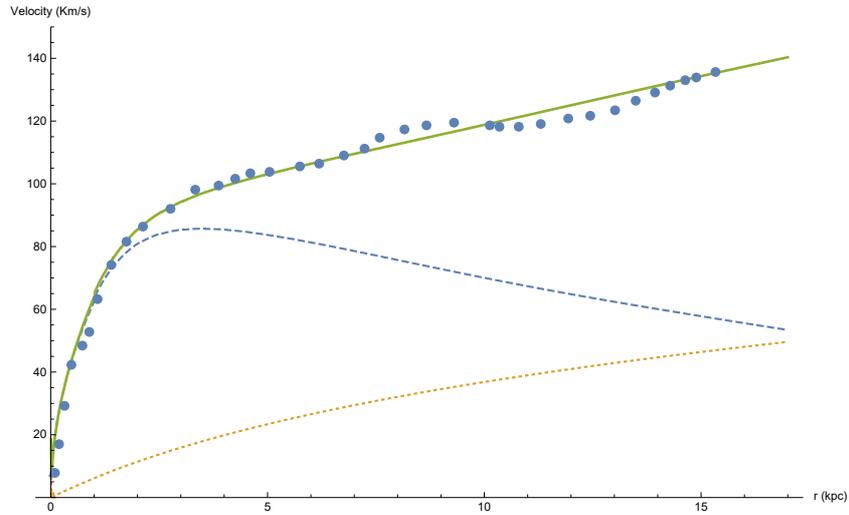}
 \caption{Velocity curve for M33. Observational points were obtained by Dr. Michal Wagman, a~former PhD student under~my supervision, using~\cite{Corbelli2}; the full line describes the  rotation curve, which is the sum of the dotted line, describing the retardation contribution, and~the dashed line, which is Newtonian.}
 \label{vcrhoc2}
\end{figure}

It is known that GR is verified by many types of observations.
However, currently, Newton–Einstein gravitational theory is at a crossroads. It has much in its favor observationally, and it has some very disquieting challenges. The~successes that it has achieved in both astrophysical and cosmological scales have to be considered in light of the fact that GR needs to appeal to two unconfirmed ingredients, dark matter and energy, to achieve these successes. Dark matter has not only been with us since the 1920s (when it was initially known as the missing mass problem), but~it has also become severe as more and more of it had to be introduced on larger distance scales as new data have become available.  Moreover, 40-year-underground and accelerator searches and experiments have failed to establish its existence. The~dark matter situation has become even more disturbing in recent years as the Large Hadron Collider was unable to find any super symmetric particle candidates, the community's preferred form of dark matter.
While things may still take turn in favor of the dark matter hypothesis, the~current situation is serious enough to consider the possibility that the popular paradigm might need to be amended in some way if not replaced altogether. In our recent work we have sought such a modification. Unlike other ideas such as Milgrom's MOND \cite{Mond}, Mannheim's Conformal Gravity \cite{Mannheim0,Mannheim1,Mannheim2},
Moffat's MOG \cite{MOG} or $f(R)$ theories and scalar-tensor gravity \cite{Corda}, the~present approach is, the~minimalist one adhering to the razor of Occam. It suggests to replace dark matter by the retardation effect within standard GR.
 Fritz Zwicky noticed in 1933 that the velocities of Galaxies within the Comma Cluster are much higher than those predicted by the virial calculation that assumes Newtonian theory~\cite{zwicky}.  He~calculated that the amount of matter required to
  account for the velocities could be 400 times greater with respect to that of visible matter, which led to suggesting  dark
   matter throughout the cluster. In~1959, Volders, pointed out  that  stars in the outer rims of the nearby galaxy M33
   do not move according to Newtonian theory~\cite{volders}. The virial theorem coupled with Newtonian Gravity implies that $MG/r \sim M v^2$, thus the~expected rotation curve should at some point decrease as $1/\sqrt{r}$.
   During the seventies Rubin and Ford~\cite{rubin1,rubin2} showed that, for~a large number of spiral galaxies, this behavior can be considered generic: velocities at the rim of the galaxies do not decrease— but they attain a plateau at some unique velocity, which differs for every galaxy. We have shown that such velocity curves can be deduced from GR if retardation is not neglected. The~derivation of the retardation force is described in previous publications~\cite{YaRe1,ge,YaRe2,Wagman,YaRe3,YaRe4,YaRe5,YaRe6}, see for example figure \ref{vcrhoc2}. It should be noted that "dark matter" effects on light rays (gravitational lensing) are well handled when taking into account retardation \cite{YaRe7}.
\begin{figure}
\centering
\includegraphics[width= 11 cm]{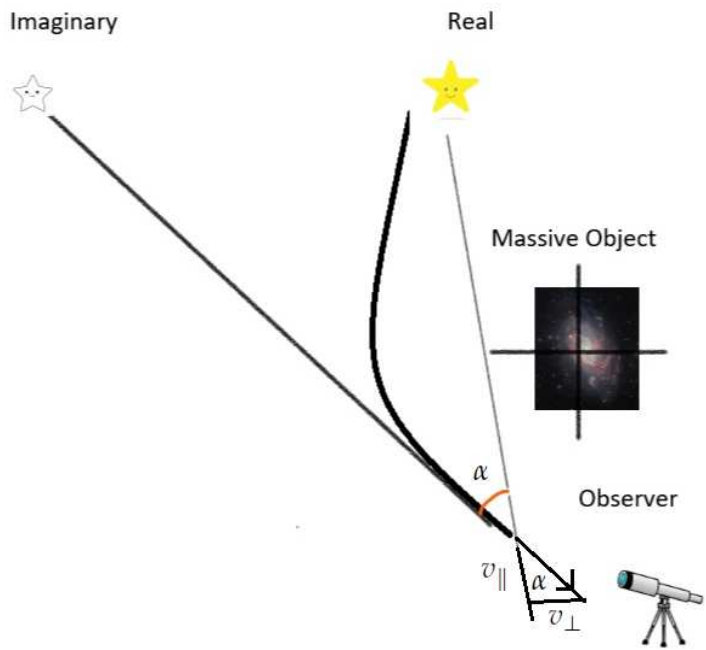}
 \caption{Light travelling toward the observer is bent due to the gravitational field of a massive object, thus a distant star appears to the observer at an angle $\alpha$ with resect to its true location. }
 \label{lensifig}
\end{figure}
The effects of gravitational lensing and the explanation of he anomalous perihelion advance of the planet Mercury where the first corroborated predictions of GR. Einstein made unpublished work on gravitational lensing as early as 1912 \cite{[4]} (see figure \ref{lensifig}). As we mentioned previously in 1915 Einstein showed how GR explained the anomalous perihelion advance of the planet Mercury without any arbitrary parameters \cite{[11]} (for a detailed account of Einstein's previous unsuccessful attempts to obtain the same see Weinstein \cite{Weinstein}) , in 1919 an expedition led by Eddington confirmed GR's prediction for the deflection of starlight by the Sun in the total solar eclipse of May 29, 1919 \cite{[12],[12a]}, making Einstein famous \cite{[11]} instantly. The reader should recall that there was a special significance to a British scientist confirming the prediction of a German scientist after the bloody battles of world war I.

Gravitational retardation effects do not seem to be very important in the solar system, up to small
corrections the dynamics is well described by Newtonian mechanics, and even the small GR effects observed can be well explained by a constant Schwarzschild metric, that is without time dependency effects of the metric.
Still, it is perhaps too early to dismiss any time dependent effects on account of the discrepancy described in table \ref{prehelexth} with regard to the perihelion precession of Mercury which is the reason for our study.

\section {General Relativity}

The general theory of relativity is based on two fundamental equations, the~first being Einstein equations~\cite{Narlikar,Weinberg,MTW,Edd}:
\beq
G_\mn = -\frac{8 \pi G}{c^4} T_\mn
\label{ein}
\enq
$G_\mn$ stands for the Einstein tensor (see \ern{Eint}), $T_\mn$ indicates the
stress–energy tensor  (see \ern{fltens}), (Greek letters are indices in the range $0-3$).
The second fundamental equation that GR is based on is the geodesic equation:
\beq
\frac{d^2 x^\alpha}{dp^2}+\Gamma^\alpha_\mn \frac{d x^\mu}{dp} \frac{d x^\nu}{dp} = 0
\label{geo}
\enq
$x^\alpha (p)$ are the coordinates of the particle in spacetime, $p$ is a typical parameter along the trajectory that for massive particles is chosen to be the length of the trajectory ($p=s$), $\tilde u^\mu = \frac{d x^\mu }{d s}$  or the proper time ($p=\tau=\frac{s}{c}$) $ u^\mu = \frac{d x^\mu }{d \tau}=c \tilde u^\mu$  is the $\mu$-th component of the 4-velocity of a massive particle moving along the geodesic trajectory  (increment of $x$ per $p$) and $\Gamma^\alpha_\mn$ is the affine connection (Einstein summation convention is assumed). The~stress–energy tensor of matter is usually taken in the form:
\beq
T_\mn = (pr+\rho c^2) \tilde u_\mu  \tilde u_\nu - pr \ g_\mn
\label{fltens}
\enq
In the above, $pr$ is the pressure and $\rho$ is the {\bf mass} density. We remind the reader that lowering and raising
indices is done through the metric $g_\mn$ and  inverse metric $g^\mn$, such that $u_\mu= g_\mn u^\nu$. The~same metric serves to calculate $s$:
\beq
ds^2 = g_\mn dx^\mu  dx^\nu,
\label{intervale}
\enq
The affine connection is connected to the metric as follows:
\beq
\Gamma^\alpha_\mn = \frac{1}{2} g^\ab \left(g_{\beta \mu, \nu} + g_{\beta \nu, \mu} - g_{\mn, \beta}\right), \qquad
 g_{\beta \mu, \nu} \equiv \frac{\partial g_{\beta \mu}}{\partial x^\nu}
\label{affine}
\enq
Using the affine connection we calculate the Riemann and Ricci tensors and the curvature scalar:
\beq
R^\mu_{\nu \ab} = \Gamma^\mu_{\nu \alpha,\beta} - \Gamma^\mu_{\nu \beta,\alpha} + \Gamma^\sigma_{\nu \alpha} \Gamma^\mu_{\sigma \beta}
- \Gamma^\sigma_{\nu \beta} \Gamma^\mu_{\sigma \alpha}, \quad R_{\ab}= R^\mu_{\ab \mu}, \quad R= g^\ab R_{\ab}
\label{RieRicci}
\enq
which, in~turn, serves to calculate the Einstein tensor:
\beq
G_{\ab}= R_{\ab} - \frac{1}{2} g_\ab R.
\label{Eint}
\enq
Hence, the~given matter distribution determines the metric through Equation~(\ref{ein}) and the metric determines the geodesic trajectories through Equation~(\ref{geo}).

\section {Linear Approximation of GR - Justification}

Only in extreme cases of compact objects (black holes and neutron stars) and the primordial reality or the very early universe does one need not consider the solution of the full non-linear Einstein Equation~\cite{YaRe1}. In~typical cases of astronomical interest\footnote{Private communication with the late Professor Donald Lynden-Bell} (certainly everywhere in the solar system) one can use a linear approximation to those equations around the flat Lorentz metric $\eta_{\mn}$ such that :
 \beq
 g_{\mn} = \eta_{\mn} + h_{\mn}, \quad \eta_{\mn} \equiv \ {\rm diag } \ (1,-1,-1,-1),
 \qquad
 |h_{\mn}|\ll 1.
 \label{lg}
 \enq
 \beq
ds^2 = (\eta_{\mn} + h_{\mn}) dx^\mu  dx^\nu,
\label{intervale1b}
\enq
 In order to appreciate the above statements, let us look at the Schwarzschild metric \cite{Weinberg}. This metric describe a static spherically symmetric mass distribution and thus is less general than the approach we intend to develop in this paper. It does have one advantage, however, and this is the ability to take into account strong gravitational fields and not
just weak ones. This advantage is irrelevant in most astronomical cases in which gravity is weak and must be only considered for trajectories near compact objects (black holes \& neutron stars).
Here we introduce it just for the sake of making an order of magnitude estimate.
The Schwarzschild squared interval can be written as:
\beq
ds_{Schwarzschild}^2=\left(1-{\frac {r_s}{r'}}\right)c^{2} dt^{2}-\left(1-{\frac {r_s}{r'}}\right)^{-1}dr'^{2}
-r'^{2}\left(d\theta ^{2}+\sin ^{2}\theta \,d\Phi ^{2}\right)
\label{Schwa}
\enq
In which $r',\theta,\Phi$ are  spherical coordinates and the point massive body
is located at $r'=0$. The Schwarzschild radius is defined as:
\beq
r_s = \frac{2 G M}{c^2}
\label{Schwaradius}
\enq
in which $M$ is the mass of the point particle. The most massive object in the solar system is
the sun with a solar mass of $M_{sun} \simeq 1.99 \ 10^{30} \ {\rm kg} $ leading to $r_s \simeq 2950 \ {\rm m}\simeq 3 \ {\rm km}$. The deviation of the metric from the empty space Minkowski metric
according to \ern{Schwa} is :
\beq
h_{00} = - \frac{r_s}{r'}.
\label{Schwaradius2}
\enq
$h_{00}$ is strongest on the sun's surface where $r' = 6.96 \ 10^8 \ {\rm m} $
in which $h_{00} \simeq 4.27 \ 10^{-6}$, this is quite small (with respect to unity) indeed. For Mercury the closer distance to the sun (perihelion) is $r' = 4.6 \ 10^{10} \ {\rm m} $ hence
$h_{00} \simeq 6.41 \ 10^{-8}$ at most. Neglecting second order terms thus means neglecting
terms of order $10^{-15}$ and seems quite justified.

\section {Linear Approximation of GR - The Metric}
\label{linear}

 One then defines the quantity:
 \beq
 \bar h_\mn \equiv h_\mn -  \frac{1}{2} \eta_\mn h, \quad h = \eta^{\mn} h_{\mn},
 \label{bh}
 \enq
 $\bar h_\mn = h_\mn $ for non diagonal terms. For~diagonal terms:
 \beq
 \bar h = - h \Rightarrow  h_\mn = \bar h_\mn -  \frac{1}{2} \eta_\mn \bar h .
 \label{bh2}
 \enq
   It~can be shown (\cite{Narlikar} page 75, exercise 37, see also~\cite{Edd,Weinberg,MTW}) that one can  choose a gauge such that the Einstein equations are:
 \beq
\bh_{\mn, \alpha}{}^{\alpha}=-\frac{16 \pi G}{c^4} T_\mn , \qquad \bh_{\mu \alpha,}{}^{\alpha}=0.
\label{lineq1}
\enq
\Ern{lineq1} can always be integrated to take the form~\cite{Jackson}:
 \ber
& & \bh_{\mn}(\vec x, t) = -\frac{4 G}{c^4} \int \frac{T_\mn (\vec x', t-\frac{R}{c})}{R} d^3 x',
\nonumber \\
 t &\equiv& \frac{x^0}{c}, \quad \vec x \equiv x^a \quad a,b \in [1,2,3], \nonumber \\
  \vec R &\equiv& \vec x - \vec x', \quad R= |\vec R |.
\label{bhint}
\enr
For reasons why the symmetry between space and time is broken, see~\cite{Yahalom,Yahalomb}.
The factor before the integral is small: $\frac{4 G}{c^4} \simeq 3.3 \times 10^{-44}$ in MKS units; hence, in~the above calculation one can take $T_\mn$, which is zero order in $h_\abp$.
In~the zeroth order:
\ber
\tilde u^0 &=& \frac{1}{\sqrt{1-\frac{v^2}{c^2}}} \equiv \gamma, \quad \vec{\tilde u} \equiv
(\tilde u^1,\tilde u^2,\tilde u^3) = \frac{\frac{\vec v}{c}}{\sqrt{1-\frac{v^2}{c^2}}}
= \vec \beta \gamma ,
\nonumber \\
v^\mu &\equiv&  \frac{d x^\mu}{d t}, \quad \vec v \equiv  \frac{d \vec x}{d t}, \quad v= |\vec v|, \quad \vec \beta  \equiv \frac{\vec v}{c}, \quad \beta = |\vec \beta|  .
\label{uz}
\enr
And also:
\beq
u^0=c \gamma, \quad \vec{u} \equiv
(u^1,u^2, u^3) = \vec v \gamma .
\label{uz2}
\enq
Assuming the reasonable assumption that the said massive body is composed of particles of non relativistic velocities:
\beq
\tilde u^0 \simeq 1,  \qquad \vec{\tilde u} \simeq \vec \beta , \qquad \tilde u^a \ll \tilde u^0   \qquad {\rm for} \quad v \ll c.
\label{uzslo}
\enq
and also:
\beq
 u^0 \simeq c,  \qquad \vec{ u} \simeq \vec v , \qquad  u^a \ll u^0   \qquad {\rm for} \quad v \ll c.
\label{uzslo2}
\enq

Let us now look at equation~(\ref{fltens}). We~assume $\rho c^2 \gg pr$ and, taking into account equation~(\ref{uzslo}), we~arrive at $T_{00} = \rho c^2 $, while  other tensor components are significantly smaller. Thus, $\bar h_{00}$ is significantly larger than other components of $\bar h_\mn$ which are ignored from now on. One should notice that it is  possible to deduce from the  gauge condition in equation~(\ref{lineq1}) the relative order of magnitude of the relative components of $h_\mn$:
\beq
\bar h_{\alpha 0,}{}^0=-\bar h_{\alpha a,}{}^a \qquad \Rightarrow
\bar h_{00,}{}^0=-\bar h_{0 a,}{}^a, \quad \bar h_{b0,}{}^0=-\bar h_{b a,}{}^a.
\label{gaugeim}
\enq

Thus, the~zeroth derivative of $\bar h_{00}$ (which contains a $\frac{1}{c}$ as $x^0 = c t$) is the same order as the spatial derivative
of $\bar h_{0a}$  meaning that $\bar h_{0a}$ is of order $\frac{v}{c}$ smaller than $\bar h_{00}$.
And the zeroth derivative of $\bar h_{0a}$ (which appears in Equation~(\ref{gaugeim})) is the same order as the spatial derivative of $\bar h_{ab}$. Meaning that $\bar h_{ab}$ is of order $\frac{v}{c}$ with respect to $\bar h_{0a}$ and of order $(\frac{v}{c})^2$ with respect to $\bar h_{00}$.

In the current approximation, the following results hold:
 \beq
 \bar h = \eta^\mn \bar h_\mn = \bar h_{00}.
 \label{bh5}
 \enq
 \beq
 h_{00} = \bar h_{00} -  \frac{1}{2} \eta_{00} \bar h = \frac{1}{2} \bar h_{00} .
 \label{bh6}
 \enq
 \beq
 h_{\underline{aa}} = -  \frac{1}{2} \eta_{\underline{aa}} \bar h = \frac{1}{2} \bar h_{00} .
 \label{bh7}
 \enq
 (The underline $\underline{aa}$ signifies that the Einstein summation convention is not assumed).
\beq
 h_{\mn} = \bar h_{\mn} = 0, \qquad \mu \neq \nu .
 \label{bh8}
 \enq
We can summarize the above results in a concise formula:
\beq
 h_{\mn} = h_{00} \delta_{\mn}.
 \label{bh9}
  \enq
in which $\delta_{\mn}$ is Kronecker's delta. Thus:
\beq
ds^2 = (\eta_{\mn} + h_{\mn}) dx^\mu  dx^\nu = (\eta_{\mn} +  h_{00} \delta_{\mn}) dx^\mu  dx^\nu = (1 +  h_{00}) c^2 dt^2 - (1 -  h_{00}) d\vec x^2 ,
\label{intervale3b}
\enq
and the proper time is:
\beq
d\tau = dt \sqrt{ (1 +  h_{00}) - (1 -  h_{00}) \beta^2}
\simeq dt \left[\sqrt{1 - \beta^2} + \frac{1}{2}h_{00} \frac{1 + \beta^2}{\sqrt{1 - \beta^2}}  \right],
\label{propt}
\enq
for slow particles this reduces to:
\beq
d\tau =  dt \left[1 + \frac{1}{2}h_{00}  \right],
\label{propt2}
\enq
and thus:
\beq
\vec u  = \frac{d \vec x}{d \tau} = \frac{d \vec x}{d t} \frac{d t}{d \tau}
=  \frac{\vec v}{\sqrt{1 - \beta^2} + \frac{1}{2}h_{00} \frac{1 + \beta^2}{\sqrt{1 - \beta^2}} }.
\label{uvrel}
\enq
The above somewhat complex relation is the main reason that we will prefer to perform our analysis
with $\vec u$ rather than $\vec v$ keeping in mind that in the solar system $\vec u \simeq \vec v$ for slow moving bodies (with respect to the speed of light).

It will be useful to introduce the gravitational potential $\phi$  which is defined below and can be calculated using Equation~(\ref{bhint}):
\beq
\phi \equiv \frac{c^2}{4} \bar h_{00}
= -\frac{ G}{c^2} \int \frac{T_{00} (\vec x', t-\frac{R}{c})}{R} d^3 x'
= -G \int \frac{\rho (\vec x', t-\frac{R}{c})}{R} d^3 x'
\label{phi}
\enq
from the above definition and \ern{bh6} it follows that:
\beq
 h_{00} = \frac{2}{c^2} \phi, \qquad \phi  = \frac{c^2}{2} h_{00}.
 \label{phih00}
  \enq
Let us now calculate the affine connection in the linear approximation:
\beq
\Gamma^\alpha_\mn = \frac{1}{2} \eta^\abp \left(h_{\beta \mu, \nu} + h_{\beta \nu, \mu} - h_{\mn, \beta}\right).
\label{affinel}
\enq
taking into account \ern{bh9} we arrive at the result:
\beq
\Gamma^a_\mn = \frac{1}{2} \left( \eta^{a \mu}  h_{00, \nu} + \eta^{a \nu}  h_{00, \mu} +  h_{00, a} \delta_\mn \right).
\label{affinel2}
\enq
the above equation is only correct for a Latin index $a$, this is our main concern, as we will concentrate on the analysis of $u^a$ (rather than $u^0$).

\section {Linear Approximation of GR - The Trajectory}

Let us start calculating the trajectory by inserting \ern{affinel2} into \ern{geo},
we arrive at the equation:
\beq
\frac{d u^a}{d\tau}= - \Gamma^a_\mn u^\mu u^\nu = u^a u^\nu h_{00, \nu}  - \frac{1}{2}
u^\nu u^\nu  h_{00, a}
\label{geo2}
\enq
We may write:
\beq
u^\nu h_{00, \nu} = h_{00, \nu} \frac{d x^\nu}{d \tau}  = \frac{d h_{00}}{d \tau}.
\label{geo3}
\enq
Thus:
\beq
\frac{d u^a}{d\tau}= u^a \frac{d h_{00}}{d \tau}  - \frac{1}{2} u^\nu u^\nu  h_{00, a}.
\label{geo4}
\enq
As the current analysis is only valid to first order in $h_{00}$ and since the right hand
side of the equation is already linea in $h_{00}$ we only need to consider the other expressions
in the right hand side to zeroth order in $h_{mn}$, thus:
\beq
\frac{d u^a}{d\tau}=
 u^{(0)a} \frac{d h_{00}}{d \tau}  - \frac{1}{2} u^{(0)\nu} u^{(0)\nu}  h_{00, a}.
\label{geo5}
\enq
It follows that according to \ern{intervale}:
\beq
c^2 = \frac{ds^2}{d\tau^2} = \eta_\mn u^{(0)\mu} u^{(0)\nu}= ( u^{(0)0})^2 - u^{(0)a} u^{(0)a}
\label{intervale2}
\enq
and also:
\beq
u^{(0)\nu} u^{(0)\nu}= ( u^{(0)0})^2 + u^{(0)a} u^{(0)a} = 2 ( u^{(0)0})^2 - c^2.
\label{intervale3}
\enq
Thus we obtain:
\beq
\frac{d u^a}{d\tau}=
 u^{(0)a} \frac{d h_{00}}{d \tau} + (\frac{1}{2}  c^2 - (u^{(0)0})^2)  h_{00, a},
\label{geo6}
\enq
or in vector form:
\beq
\frac{d \vec u}{d\tau}=
 {\vec u}^{(0)} \frac{d h_{00}}{d \tau} + (\frac{1}{2}  c^2 - (u^{(0)0})^2) \vec \nabla h_{00},
\label{geo7}
\enq
In term of the potential $\phi$ this takes the form:
\beq
\frac{d \vec u}{d\tau}=
 \frac{2\vec u^{(0)}}{c^2} \frac{d \phi}{d \tau} + (1 - \frac{2(u^{(0)0})^2}{c^2} ) \vec \nabla \phi.
\label{geo8b}
\enq
If the body is slow moving such that $\beta \rightarrow 0$ it follows that $\tau \simeq t$, 
$\vec u \simeq \vec v$ , $u^{(0)0} \simeq c$ and thus:
\beq
\frac{d \vec v}{dt}=  \frac{2\vec v^{(0)}}{c^2} \frac{d \phi}{d t} - \vec \nabla \phi.
\label{geo8c}
\enq
if $\phi$ does not depend explicitly on time: $\frac{d \phi}{d t} = \vec v \cdot \vec \nabla \phi$
and we obtain:
\beq
\frac{d \vec v}{dt}=  2 \vec \beta (\vec \beta \cdot \vec \nabla \phi) - \vec \nabla \phi
\simeq - \vec \nabla \phi.
\label{geo8d}
\enq
thus we are back to the Newtonian equation of motion, which can be used as a good approximation for most purposes to derive planetary motion.

\subsection {"Angular Momentum"}
\label{Angmom}

Let us perform a vector multiplication of \ern{geo7} with the vector $\vec r = (x^1,x^2,x^3)$.
We obtain:
\beq
\vec r \times \frac{d \vec u}{d\tau}=
 \vec r \times {\vec u}^{(0)} \frac{d h_{00}}{d \tau} + (\frac{1}{2}  c^2 - (u^{(0)0})^2) \vec r \times \vec \nabla h_{00},
\label{geo8}
\enq
Now suppose that $h_{00} = h_{00} (r)$ in which $r = |\vec r|$, in this case $\vec r \times \vec \nabla h_{00} = 0$. And thus:
\beq
\frac{d (\vec r \times \vec u)}{d\tau}= \vec r \times \frac{d \vec u}{d\tau}=
 \vec r \times {\vec u}^{(0)} \frac{d h_{00}}{d \tau} = (\vec r \times \vec u) \frac{d h_{00}}{d \tau}
\label{geo9}
\enq
 the last equation sign is correct to first order in $h_{00}$ (which is the order of our entire analysis). We now define an "angular momentum":
 \beq
\vec J \equiv m \vec r \times \vec u \simeq m \vec r \times \vec v
\label{JV}
\enq
in which $m$ is the mass of the particle under consideration. The right hand side of the
equation is correct for slow bodies, in the solar system. It follows that:
\beq
\frac{d \vec J }{d\tau}= \vec J \ \frac{d h_{00}}{d \tau}
\label{Jeq1}
\enq
Hence for each cartesian component:
\beq
\frac{d J_x }{d\tau}= J_x \ \frac{d h_{00}}{d \tau}, \quad
\frac{d J_y }{d\tau}= J_y \ \frac{d h_{00}}{d \tau}, \quad
\frac{d J_z}{d\tau}= J_z \ \frac{d h_{00}}{d \tau}.
\label{Jeq2}
\enq
Or also:
\beq
\frac{d (\ln J_x - h_{00}) }{d\tau}= 0, \quad
\frac{d (\ln J_y - h_{00}) }{d\tau}= 0, \quad
\frac{d (\ln J_z - h_{00}) }{d\tau}= 0, \quad.
\label{Jeq3}
\enq
Thus:
\beq
\vec J = \vec J_0 e^{h_{00}} \simeq \vec J_0 (1+ h_{00})
\label{Jeq4}
\enq
in which $\vec J_0$ is constant. This implies that the "angular momentum" has a size that depends on $h_{00}$:
\beq
J \equiv |\vec J| = |\vec J_0| e^{h_{00}}  \equiv J_0 e^{h_{00}}
\label{Jeq5}
\enq
and a fixed direction:
\beq
\hat J \equiv \frac{\vec J}{J} = \frac{\vec J_0}{J_0} \equiv \hat J_0.
\label{Jeq6}
\enq
We conveniently choose this direction to point in the $z$ axis, hence:
\beq
\vec J = \vec J_0 e^{h_{00}} = J_0 e^{h_{00}} \hat z, \qquad J_x=J_y=0.
\label{Jeq7}
\enq
Since the direction of $\vec J$ is fixed and perpendicular to the direction of both $\vec r$ and  $\vec u$ it follows that both vectors lie in the $x-y$ plane. Thus we can conveniently describe their motion using the standard polar coordinates $r,\theta$. We underline that unlike the angular momentum vector of classical mechanics this vector is only fixed in direction but not in size. We also underline that nowhere did we imply that the velocity of the particle must be small with respect to the velocity of light.

\subsection {"Energy"}
\label{Ener}

Let us make a scalar multiplication of \ern{geo7} with $\vec u$ to obtain:
\beq
\vec u \cdot \frac{d \vec u}{d\tau}=
 {\vec u}^2 \frac{d h_{00}}{d \tau} + (\frac{1}{2}  c^2 - (u^{0})^2) \vec u \cdot  \vec \nabla h_{00},
\label{energy1}
\enq
For a $h_{00}$ without explicit time dependence we have:
\beq
\frac{d h_{00}}{d \tau} = \vec u \cdot  \vec \nabla h_{00}+ \tilde u^0 \frac{\partial h_{00}}{\partial t}
= \vec u \cdot  \vec \nabla h_{00},
\label{energy2}
\enq
it follows that:
\beq
\frac{1}{2} \frac{d {\vec u}^2}{d\tau}=
(\frac{1}{2}  c^2  + {\vec u}^2  - (u^{0})^2) \frac{d h_{00}}{d \tau},
\label{energy3}
\enq
the term in parenthesis in the right hand side needs only to be evaluated to zeroth order
in $h_{00}$ and this can be done using \ern{intervale2}:
\beq
\frac{1}{2} \frac{d {\vec u}^2}{d\tau}=
(\frac{1}{2}  c^2  - c^2) \frac{d h_{00}}{d \tau} = -\frac{1}{2}  c^2 \frac{d h_{00}}{d \tau}
= -\frac{d \phi}{d \tau},
\label{energy4}
\enq
in which we use the definition for the potential given in \ern{phih00}. Hence we obtain the conserved "energy":
\beq
E \equiv \frac{1}{2} m {\vec u}^2 + m\phi \simeq \frac{1}{2}m v^2 + m \phi.
\label{energy5}
\enq
the "energy" is defined for either fast or slow bodies, while the right hand $\simeq $ relates only to slow bodies.

\subsection {Polar coordinates}

Introducing polar coordinates we may write \ern{JV} and \ern{energy5} as follows:
\beq
E = \frac{1}{2} m  \left[\dot r ^2  + r^2 \dot \theta^2\right] + m\phi,
\qquad J = m r^2 \dot \theta, \qquad \dot r \equiv \frac{dr}{d \tau}, \quad
\dot \theta \equiv \frac{d\theta}{d \tau}
\label{energyJ}
\enq
 the above forms seem quite classical, however, one should remember that the derivatives are with respect to the proper time of the body and not with respect to the global $t$ coordinate. One should also recall that $J$ is not constant and can vary to a small extent. Thus we eliminate $\dot \theta$ and obtain:
 \beq
E = \frac{1}{2} m  \dot r ^2  + \frac{J^2}{2 m r^2} + m\phi
= \frac{1}{2} m  \dot r ^2  + \frac{J_0^2 e^{2 h_{00}}}{2 m r^2} + m\phi
\simeq \frac{1}{2} m  \dot r ^2  + \frac{J_0^2}{2 m r^2} + m\phi + \frac{2 J_0^2 \phi }{m c^2 r^2}.
\label{energyJ2}
\enq
The above result is quite classical except the last term which is a relativistic correction as
is disclosed by the $\frac{1}{c^2}$ appearing there.
To put the above result in a form which is closer to the forms which are found in the literature
\cite{padma,MTW}
we make the following observation. We have
written $d\vec x^2 = dr^2 + r^2 \left(d\theta ^{2}+\sin^2 \theta d\Phi ^{2}\right)$ as is usually done for spherical coordinates. But notice that $r$ in the above is not strictly a radial coordinate which is defined as the circumference, divided by $2 \pi$, of a sphere centered around the massive body. In fact from \ern{intervale3b} it is clear that the appropriate radial coordinate is:
\beq
r' = r \sqrt{1-  h_{00}}
\label{rp}
\enq
which is a small correction to $r$. Now calculating the differential $dr'$ it follows that:
\beq
dr' = dr \sqrt{1-  h_{00}} + r d \sqrt{1-  h_{00}} = dr \frac{1-h_{00} - \frac{1}{2} r \frac{d h_{00}}{dr}}{\sqrt{1- h_{00}}}.
\label{drp}
\enq
If we assume a Schwarzschild metric according to \ern{Schwaradius2}:
\beq
h_{00}= - \frac{r_s}{r'} =- \frac{r_s}{r \sqrt{1-  h_{00}} }
\Rightarrow  h_{00} \sqrt{1-  h_{00}} = - \frac{r_s}{r} \Rightarrow  h_{00} = - \frac{r_s}{r},
\qquad r' \simeq r \gg r_s
\label{hSchwa3}
\enq
where the last equality is correct to first order. Alternatively one can use \ern{phi} to calculate the gravitational potential for a static point mass to obtain:
\beq
\phi = -G \frac{M}{r}
\label{phipoin}
\enq
and then plug this into \ern{phih00} to obtain again:
\beq
 h_{00} = \frac{2}{c^2} \phi = -\frac{2 G M}{c^2 r} = -\frac{r_s}{r}.
 \label{phih00poi}
  \enq
It now follows that:
\beq
\frac{d h_{00}}{dr} = \frac{r_s}{r^2} = - \frac{h_{00}}{r}
\label{dh00poi}
  \enq
Plugging \ern{dh00poi} into \ern{drp} leads to:
\beq
dr' = dr \frac{1-h_{00} - \frac{1}{2} r \frac{d h_{00}}{dr}}{\sqrt{1- h_{00}}}
= dr \frac{1- \frac{1}{2}h_{00}}{\sqrt{1- h_{00}}}.
\label{drp2}
\enq
Hence to first order in $h_{00}$:
\beq
dr' = dr.
\label{drp3}
\enq
Using the results \ern{rp} and \ern{drp3} the interval given in \ern{intervale3b} can be rewritten as:
\ber
ds^2 &=& (1 +  h_{00}) c^2 dt^2 - (1 -  h_{00}) d\vec x^2
\nonumber \\
&=& (1 +  h_{00}) c^2 dt^2 - (1 -  h_{00}) dr'^2 - r'^2 \left(d\theta ^{2}+\sin^2 \theta d\phi ^{2}\right).
\label{intervale4b}
\enr
As to first order in $h_{00}$:
\beq
 1 -  h_{00} = \frac{1}{1 +  h_{00}},
\label{intervale5b}
\enq
and taking into account \ern{Schwaradius2} we obtain:
\beq
ds^2 = (1 - \frac{r_s}{r'}) c^2 dt^2 - (1 -\frac{r_s}{r'})^{-1}  dr'^2 - r'^2 \left(d\theta ^{2}+\sin^2 \theta d\phi ^{2}\right) = ds_{Schwarzschild}^2 .
\label{intervale6b}
\enq
Thus to first order our metric is identical to Schwarzschild's for the case of a static point particle. This makes our analysis superior as it addresses the case of a general density distribution and does not ignore the possibility of time dependence which is crucial for
retardation effects to take place.

In terms of $r'$ we may write the energy as:
\ber
& & \hspace{-1 cm} E = \frac{1}{2} m  \dot r'^2  + \frac{J_0^2 e^{2 h_{00}}}{2 m r'^2} (1-h_{00}) + m\phi
= \frac{1}{2} m  \dot r'^2  + \frac{J_0^2}{2 m r'^2}(1+h_{00}) + m\phi
\nonumber \\
&\hspace{-2 cm}=& \hspace{-1 cm} \frac{1}{2} m  \dot r'^2  + \frac{J_0^2}{2 m r'^2} + m\phi + \frac{ J_0^2 \phi }{m c^2 r'^2}
= \frac{1}{2} m  \dot r'^2  + \frac{J_0^2}{2 m r'^2} - \frac{G M m}{r'} -
 \frac{G M J_0^2 \phi }{m c^2 r'^3}.
\label{energyJ3}
\enr
The relativistic correction has a $\frac{1}{r'^3}$ dependence and the is greater the more the object is close to the gravitating point mass which represents the sun. We also notice that in terms of $r'$ one may write up to first order in $h_{00}$:
\beq
 J_0 = m r'^2 \dot \theta
\label{Jrp}
\enq
which is reminiscent of the well known classical result regrading the conservation of angular momentum.

\subsection {Integration of the equations of motion}

Using \ern{Jrp} we may write:
\beq
\dot r' = \dot \theta \frac{d r'}{d \theta} = \frac{J_0}{m r'^2} \frac{d r'}{d \theta}
\label{rpd}
\enq
Defining\footnote{The symbol $u'$ has nothing to do with the previously defined $u^\nu$ and is chosen this way in order to comply with the symbol used in the literature \cite{padma}.}:
\beq
u' \equiv \frac{1}{r'} \Rightarrow \dot r' = - \frac{J_0}{m} \frac{d u'}{d \theta}
\label{up}
\enq
Plugging \ern{up} into \ern{energyJ3} we arrive at the result:
\beq
\tilde E \equiv \frac{2 m E}{J_0^2}  = \left(\frac{d u'}{d \theta} \right)^2
+u'^2 - \frac{2 G M m^2}{J_0^2} u' - \frac{2 G M }{c^2} u'^3
\label{tildE}
\enq
Or:
\beq
\left(\frac{d u'}{d \theta} \right)^2 = \tilde E - u'^2 + \frac{2 G M m^2}{J_0^2} u'
+ \frac{2 G M }{c^2} u'^3
\label{tildE2}
\enq
Which is exactly Padmanabhan \cite{padma} equation 7.114 (page 318), with a slightly different notation:
\beq
L_{Padmanabhan} = J_0, \qquad \varepsilon_{Padmanabhan} = m c^2 \sqrt{1+ 2 \frac{E}{m c^2}}
\label{tildE3}
\enq
The analysis of Padmanabhan \cite{padma} of \ern{tildE2} leads to \ern{prehprec} and will not be repeated here due to lack of space, the interested reader is referred to the original text.
We only mention that the semi-major axis, can be calculated by:
\beq
a = \frac{J_0^2}{G M (1-e_{eccentricity}^2)m^2}
\label{semimajoraxis}
\enq
and the eccentricity can be calculated as:
\beq
e_{eccentricity}= \sqrt{1 + \frac{2 E_N J_0^2}{G^2 M^2 m^3 }}, \qquad
E_N \equiv \varepsilon_{Padmanabhan} - mc^2.
\label{eccentricity}
\enq
Recently K{\v{r}}{\'{\i}}{\v{z}}ek \cite{Krizek2} has commented regarding the uncertainties related
the approximation needed in order to derive \ern{prehprec} from \ern{tildE2}. The integration
of \ern{tildE2} involves elliptic functions, and the derivation of \ern{prehprec} is not exact
but involves approximations to those integrals, thus errors are introduced which should be accounted for.

\section {Metric correction and the gravitational potential}

So far we have considered only a naive model in which the sun was taken to be a static point
particle in the inertial frame, this has led us from \ern{phipoin} to \ern{prehprec}. However, we can certainly do better, as it is well known that \ern{phi} can be integrated for a particle $p$ of mass $M_p$ moving in an arbitrary trajectory with velocity $\vec v_p(t)$, resulting in a Li\'{e}nard-Wiechert potential \cite{Lienard,Wiechert,Jackson}:
\beq
\phi_{LWp} =\left. -\frac{G M_p}{R_p(1-\hat R_p \cdot \vec \beta_p)}\right|_{t_{rp}}, \quad \hat R_p = \frac{\vec R_p}{R_p}
, \quad  t_{rp}  = t - \frac{R_p(t_{rp})}{c}, \quad R_p(t)=|\vec x - \vec x_p (t)|.
\label{LenWick}
\enq
The peculiar thing about this equation is that it must be evaluated at a retarded time $t_{rp}$ which is only given in terms of an implicit equation. This equation requires a knowledge of the evaluation point of the potential and the knowledge of the particles trajectory for its solution, which is highly inconvenient. The derivative of the global time coordinate $t$ with respect to the particular retardation time related to a certain particle is:
\beq
\frac{dt}{dt_{rp}}= \left. \frac{1}{1-\hat R_p \cdot \vec \beta_p}\right|_{t_{rp}} \simeq 1
\label{trp}
\enq
Which implies that for a slow moving particle withe $\beta_p \ll 1$ we arrive at the classical gravitational potential:
\beq
\phi_{LWp} \simeq \phi_p = -\frac{G M_p}{R_p(t)}
\label{phicla}
\enq
If many slow moving bodies affect a particular body, this body will move under the influence of a potential:
\beq
\phi = \sum_{p=1}^{N} \phi_p = -\sum_{p=1}^{N} \frac{G M_p}{R_p(t)}
\label{phiclaN}
\enq
this is essentially  K{\v{r}}{\'{\i}}{\v{z}}ek's \cite{Krizek} equation (7) (given in
terms of potential rather than force). In the Mercury case the most dominant body influencing its
trajectory is the sun which is much more massive than the other planets in the solar system, thus
to a zeroth order approximation we can ignore the other planets altogether. However, for more subtle
effects such as the precession of the perihelion for Mercury they cannot be ignored. Indeed most
of the precession can be attributed to the effect of the various planets as is indicate in the first row of table \ref{prehelexth} \cite{Clemence,Park}. K{\v{r}}{\'{\i}}{\v{z}}ek's \cite{Krizek} has correctly criticized the lack of any error bar to this number which must result from the uncertainty in the other planet masses and trajectories. This should be contrasted with the uncertainty attributed to the observed precession of the perihelion for Mercury in the same table.

As indicated previously relativistic corrections in \ern{LenWick} are only relevant for
fast moving bodies with a considerable $\beta$. The highest $\beta$ in the solar system
is for Mercury itself with $v \simeq 47.36 ~10^3$ km/s yielding a $\beta \simeq 1.6 ~10^{-4}$.
However, the highest velocity among the other planets affecting, Mercury's orbit is Venus, with
$v \simeq 35.0 ~10^3$ km/s yielding a $\beta \simeq 1.2 ~10^{-4}$. Thus the relativistic corrections are small corrections to the contribution the other planets make to the gravitational potential experienced by Mercury, those contributions are quite small even before a relativistic correction is applied. It thus remains to analyze the corrections to the gravitational potential of the sun.

The Schwarzschild based analysis, as well as the one presented in the previous sections assumes that the Sun is static and is located at the origin of axes. The Li\'{e}nard-Wiechert potential described
above allows us to take into account various corrections to the above simplified assumptions. The sun need not be located at the origin of axis, and it can be moving with respect to the inertial frame. The motion of the sun as well as the time gravity propagates from the Sun to Mercury all affect the gravitational potential. All those effects are small but not necessarily of the same magnitude, hence we need to evaluate there relative strengths.

Classical mechanics teaches us that a frame in which the Sun is located at the origin of axis is not inertial. A classical system which is inertial is one in which the origin of axis is located at the center of mass. This is only true in classical mechanics, however, the relativistic engine effect described in \cite{MTAY1,AY1,MTAY4,AY2,RY2,RY3,RY4} shows that the relativistic retardation phenomena may cause even the center of mass to accelerate, and thus cannot be used in strictly fixing an inertial frame.

Never the less, we shall assume for now that the center of mass does define an inertial frame and discuss the effect of the motion of the Sun with respect to it. The velocity of the Sun from the center of mass is estimated to be $v \simeq 11$ m/s \cite{Jason} yielding a $\beta \simeq 3.7 ~10^{-8}$. The difference between the potential $\phi_{LWsun}$ and the potential $\phi$ 
given in \ern{phipoin} is defined as:
\beq
\Delta \phi \equiv \phi_{LWsun} - \phi, \qquad \phi_{LWsun} = \phi + \Delta \phi
\label{phiLWsun}
\enq
The difference can be written in terms of three terms:
\ber
\Delta \phi &=& \Delta \phi_{\beta} + \Delta \phi_{tr} + \Delta \phi_{d} 
\nonumber \\
\Delta \phi_{d} &=& -G M_{sun} \left(\frac{1}{R_{sun} (t)} - \frac{1}{r}\right), \qquad 
R_{sun} (t) = |\vec x - \vec x_{sun} (t)|
\nonumber \\
\Delta \phi_{tr} &=& -G M_{sun} \left(\frac{1}{R_{sun} (t_r)} - \frac{1}{R_{sun} (t)}\right), \qquad 
t_r = t - \frac{R_{sun}(t_{r})}{c}
\nonumber \\
\Delta \phi_{\beta} &=& -G M_{sun} \left(\frac{1}{(1-\hat R_{sun}(t_r) \cdot \vec \beta_{sun}(t_r)) R_{sun} (t_r)} - \frac{1}{R_{sun} (t_r)}\right),
\label{delphi}
\enr
Those terms can be approximated to first order as follows:
\ber
\Delta \phi_{d} &\simeq & -\phi ~(\hat x \cdot \hat x_{sun}) \frac{x_{sun}}{r},\qquad
\hat x \equiv \frac{\vec x}{|\vec x|} = \frac{\vec x}{r}, \quad
\hat x_{sun} \equiv \frac{\vec x_{sun}}{|\vec x_{sun}|}
\nonumber \\
\Delta \phi_{tr} &=& -\Delta \phi_{\beta} = \phi ~(\hat x \cdot \vec \beta_{sun})
\label{delphi2}
\enr
It is apparent that $\Delta \phi_{tr}$ and $\Delta \phi_{\beta}$ cancel each other to 
first order in $\beta$ as expected, while second order corrections in $\beta$ are negligible. Thus we may write a more accurate potential due to the sun as:
\beq
 \phi_{LWsun} = \phi (1 - (\hat x \cdot \hat x_{sun}) \frac{x_{sun}}{r})
\label{phiLWsun2}
\enq
Taking into account the distance of the sun from the barycenter (see figure 3 of K{\v{r}}{\'{\i}}{\v{z}}ek \cite{Krizek}) it follows that it of the order of 
the Sun's diameter, while the exact trajectory of the Sun with respect to the solar system barycenter is quite complex. Thus:
\beq
 \frac{x_{sun}}{r} \simeq \frac{2~Sun ~Radius}{Mercury~Perihelion} 
 \simeq 2 \frac{6.96~10^8}{4.60~10^{10}} 
 \simeq 3 \%
\label{phiLWsun3}
\enq
The analytical approach described above is not useful for calculating the trajectory
in the presence of the $\Delta \phi$ potential correction. As now the potential depends on time
and thus the energy is not strictly conserved, moreover the potential depends on the polar angle and thus angular momentum is not conserved in an exact sense. Thus it seems best to integrate the equations numerically, however, this is beyond the scope of the present paper.
A crude model that can accommodate the present approach would be to replace the Sun's mass with
an effective mass: 
\beq
 \phi_{LWsun} = -\frac{G M_{eff}}{r}, \qquad M_{eff} \equiv M_{sun} (1 \pm 0.03)
\label{phiLWsun4}
\enq
Thus the anomalous perihelion shift is not the value given by \ern{prehprec} but rather by:  
\beq
\delta \theta = \frac {6 \pi G M_{eff}}{a c^{2}\left(1-e^{2}\right)},
\label{prehprec2}
\enq 
for Mercury the semi-major axis is $a \simeq 57.909~10^6$ km and the eccentricity is 
$e \simeq 0.2056$ thus:
\beq
\delta \theta = 43"/{\rm cy} \pm 1.3"/{\rm cy}.
\label{prehprec3}
\enq 
Thus the calculated value can be easily put within the error bar of the observed value
as we recall that the discrepancy is only about $0.56"$ per century. Other uncertainties may arise regarding the Newtonian contributions of the other planets to the gravitational potential \cite{Krizek}.
 
\section {Conclusions}

We have shown that one can solve the Mercury trajectory problem using the weak gravitational approximation to GR and without using the Schwarzschild metric. This is contrary to the claims
that this problem can only be solved in the framework of strong gravity \cite{padma}. In fact it
can easily be seen that the weak gravity approximation will suffice anywhere in the solar system, and Mercury is no exception to this rule.

We have presented a solution to the current discrepancy between the observed and calculated perihelion shift although the solution is crude and numerical simulations may reveal more details.

The weak field approximation takes in account retardation, however, the retardation correction
is of order $\beta^2$ as should be expected and is thus unimportant in the Mercury trajectory problem, certainly comparing to the corrections related to the motion of the center of the Sun with respect to the barycenter of the solar system. In this respect the Mercury perihelion problem is quite different than the "dark matter" problem, although the weak gravity approach can handle them both.

\end{document}